\documentclass[twocolumn,showpacs,preprintnumbers,amsmath,amssymb]{revtex4}
\usepackage{amsmath}
\usepackage{epsfig}
\usepackage{epsf}
\usepackage{array}
\usepackage{graphicx}

\newcommand{\beq}{\begin{equation}}
\newcommand{\eeq}{\end{equation}}
\newcommand{\be}{\begin{equation}}
\newcommand{\ee}{\end{equation}}
\newcommand{\beqa}{\begin{eqnarray}}
\newcommand{\eeqa}{\end{eqnarray}}
\newcommand{\bea}{\begin{eqnarray}}
\newcommand{\eea}{\end{eqnarray}}
\newcommand{\eqs}{\end{split}}
\newcommand{\bqs}{\begin{split}}

\begin{document}
%\title{Bound quasiparticle states induced by charging effects}
%\title{Thermodynamic Properties of a Double-Island Qubit}
\title{Coulomb-blockade-induced bound quasiparticle states in a double-island qubit}
\author{D.~A.~Pesin}
 \affiliation{Physics Department, University of Colorado, Boulder, CO 80309}
\author{A.~V.~Andreev}
\affiliation{Physics Department, University of Colorado, Boulder, CO 80309}

\date{\today}

\begin{abstract}
  We study the low temperature thermodynamic properties of a superconducting double-island
  qubit. For an odd number of  electrons  in  the device, the ground state corresponds to
the intrinsic quasiparticle  bound to the tunneling contact. The ground state is
separated from the continuum of excited states by a finite gap
   of order of the Josephson  energy $E_J$. The presence of the bound quasiparticle
   state results in a nonmonotonic temperature dependence of the width of the transition region between
  Coulomb blockade plateaus. The minimum width occurs at the ionization temperature of the bound state,
  $T_{i}\sim E_J/\ln(\sqrt{\Delta E_J}/\delta)$, with $\Delta$ and $\delta$ being
  respectively the superconducting gap and single particle mean level spacing in the
  island. For an even number of electrons
  in the system, we show that the Coulomb enhancement of the Josephson energy
  can  be significantly stronger than that in the case of a single grain coupled
  to a superconducting lead. If the electrostatic energy favors
  a single broken Cooper pair, the resulting quasiparticles are bound to the
  contact with an energy that is exponentially small in the inverse dimensionless
  conductance.
\end{abstract}

\pacs{73.23.Hk, 74.50.+r, 85.35.Gv}

\maketitle

Some of the most promising qubit realizations are based on superconducting
nanodevices~\cite{Bouchiat1998,Mooij1999,Nakamura1999,Bibow2002,Schoelkopfnew,Martinis2002}.
These include
charge~\cite{Bouchiat1998,Mooij1999,Nakamura1999,Bibow2002,Schoelkopfnew} and
phase~\cite{Martinis2002} qubits. A charge qubit consists of a small
superconducting island coupled by a tunneling contact either to another
superconducting grain or a bulk superconducting lead. At low temperatures quantum
fluctuations of the island charge strongly influence the properties of such
devices.

\begin{figure}
\includegraphics[keepaspectratio=true,bb=0 0 190 146]{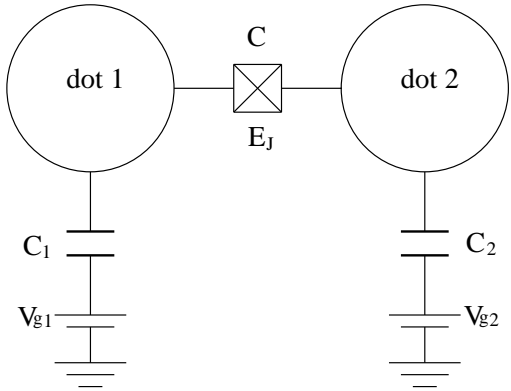}
\caption{The two islands of the qubit are coupled to the gates which are
maintained at the voltages $V_{g1}$ and $V_{g2}$.} \label{The_scheme}
\end{figure}

We study the effect of quantum charge fluctuations on the thermodynamic
properties of a double-island qubit~\cite{Bibow2002,Schoelkopfnew}. The device is
schematically depicted in Fig.~\ref{The_scheme}. It consists of two
superconducting islands coupled to each other by a tunnelling contact and
capacitively coupled to external gates. In a typical experimental realization the
probability of electron tunneling between the qubit and the surrounding
environment on the experimental time scale is negligible. Accordingly we consider
the total number of electrons in the device to be fixed~\cite{footnote}. This
number may either be even or odd and can be randomly changed by heating the
device.

In the absence of tunnelling between the grains the electron number on each
island is quantized and changes abruptly at the charge degeneracy lines in the
$V_{g1}$-$V_{g2}$ plane. Depending on the values of the pairing gap and the
charging energy electrons can be transferred between in the islands either one by
one or in pairs. At finite tunnelling between the grains quantum charge
fluctuations lead to rounding of Coulomb blockade steps in the voltage dependence
of the island charge. This phenomenon was studied
theoretically~\cite{Matveev1991,Matveev1995,Matveev1998,Schoeeler1994,Goppert1998}
and experimentally~\cite{Lehnert2003} for different single electron devices.

In this paper we determine the gate voltage dependence of the island charge for a
double-island qubit. We also study the Coulomb enhancement of the energy of
Josephson coupling between the grains. For a fixed total number of electrons in
the qubit the Hamiltonian of the system can be written as
\begin{subequations}
\label{eq:h}
\begin{eqnarray}
\label{eq:h1}
H &=& H_C+H_t+H_{0}, \\
\label{eq:h2}
H_C &=&
%E_{s}(\hat{n}+\hat{n}_2-N_s)^2+
E_C(\hat{n}-N)^2, \\
\label{eq:h3}
  H_t &=& \sum_{kp\sigma}t_{kp}a^\dagger_{k,\sigma} a_{p,\sigma} + h. c.,
\end{eqnarray}
\end{subequations}
where $t_{kp}$ are the tunneling matrix elements, $\sigma=\pm 1$
is the spin index, and $a_{p/k,\sigma}$ is the annihilation
operator for an electron in state $k,\sigma$ in the left island or
state $p,\sigma$ in the right island respectively. The operator of
the electron number difference (in units of electron charge)
between the left and right islands is denoted by $\hat{n}$, $E_C$
is the corresponding charging energy, and the dimensionless
parameter $N$ is given by the appropriate linear combination of
the voltages on the left and right gates. Finally, $H_{0}$ denotes
the Hamiltonian of the two grains in the absence of tunneling and
charging interaction. For simplicity we consider $H_0$ to be of
the BCS form and assume that both islands have identical pairing
gaps, $\Delta$, and single particle mean level spacings, $\delta$.

We assume that $\langle |t_{kp}|^2\rangle_{kp}/\delta^2 \ll 1$ , where $\langle \ldots
\rangle_{kp}$ denotes averaging over the Fermi surfaces of both
islands. Then the dimensionless conductance $g=\frac{h}{e^2 R}=8\pi^2\frac{\langle
|t_{kp}|^2\rangle_{kp}}{\delta^2}$ of
the tunneling contact,with $R$ being its
normal state resistance, obeys the condition $g \ll 8 \pi^2$.
Further, we restrict our analysis to low temperatures, $T < T^*
=\Delta/\ln(\Delta/\delta)$, where no thermal quasiparticles are
present~\cite{Tinkhambook}.

First we consider the case of an odd number of electrons in the device. To be
specific we assume it to be equal to unity and study the gate voltage dependence
of the average island charge difference $\langle \hat{n} \rangle$ for $|N| \ll
1$. In this regime the charge states with $n=\pm 1$ are nearly degenerate and
quantum charge fluctuations are strong.

In the absence of tunneling the eigenstates of the Hamiltonian, Eq.~(\ref{eq:h})
are characterized by $n$ and the momenta and spins of the quasiparticles present
in the islands. For an odd number of electrons there is precisely one
quasiparticle in the qubit at $T < T^*$. We denote the states with $n=1$ and the
quasiparticle in state $k,\sigma$ in the left island by $|k,\sigma;0\rangle $ and
those with $n=-1$ and the quasiparticle in state $p,\sigma$ in the right island
by $|0;p,\sigma \rangle $.

For weak tunneling the most straightforward way to proceed is to
use perturbation theory. In second order in tunneling scattering
matrix elements between states $|0;p,\sigma \rangle $ and
$|0;p',\sigma \rangle $, and $|k,\sigma ;0 \rangle $ and
$|k',\sigma ;0\rangle $, appear. Their magnitude $\sim
\frac{\langle |t_{kp}|^2\rangle_{kp}}{\delta}\sqrt{\frac{\Delta}{E_C N}}$ significantly
exceeds the energy distance between such states,
$\sim\delta^2/\Delta$, even far from the charge degeneracy point.
Thus in contrast to the case of a normal grain~\cite{Matveev1991}
virtual tunneling processes strongly mix different quasiparticle
states even for $\langle |t_{kp}|^2\rangle_{kp}/\delta^2 \ll 1$. For $|N| \ll 1$ only the two
nearly degenerate charge states need be considered. In the spirit
of degenerate perturbation theory~\cite{Landau} we look for the
ground state wave function in the form of a linear combination of
low energy states
\begin{equation}
\label{eq:wf} |\Psi\rangle=\sum_{p,\sigma}
C_{p,\sigma}|0;p,\sigma\rangle+\sum_{k,\sigma} C_{k,\sigma}|k,\sigma;0\rangle.
\end{equation}
We rewrite the tunneling Hamiltonian (\ref{eq:h3}) in terms of the quasiparticle
creation and annihilation operators with the aid of the Bogolubov transformation
$a_{p(k),\sigma}= u_{p(k)} \alpha_{p(k),\sigma} + \sigma v_{p(k)}
\alpha^\dagger_{-p(k),-\sigma}$ (here $u_{p(k)}$ and $v_{p(k)}$ are the standard
BCS coherence factors) and project it onto the subspace of low energy states
given by Eq.~(\ref{eq:wf}). The energy spectrum of the resulting problem can be
determined exactly. The density of states in the continuum is unaffected by the
presence of tunneling. In addition to the continuum states one bound state below
the continuum appears. Though its wave function, Eq.~(\ref{eq:wf}), depends on
the specific form of the tunneling matrix elements $t_{kp}$ its energy in the
weak tunneling approximation can be expressed in terms of the dimensionless
conductance of the tunneling contact, $g$. We obtain for the ground state energy
\begin{equation}
\label{eq:gs}
E_0(N) = E_C(1+N^2)+\Delta- \sqrt{4E_C^2N^2+E_J^2},
\end{equation}
where $E_J= g\Delta/8$ is the
Ambegaokar-Baratoff~\cite{Ambegaokar} expression for the Josephson
energy. The binding energy, i.e. the gap between the bound state
energy and the bottom of the continuum is given by
\begin{equation}\label{eq:gap}
\epsilon_B= E_J\left(-| \tilde{N} |+\sqrt{\tilde{N}^2+ 1}\right),
\end{equation}
where we introduced a rescaled gate voltage
%\begin{equation}\label{eq:N_rescaled}
    $\tilde N =\frac{2 E_C}{E_J}\,N$.
%\end{equation}

Using the obtained energy spectrum we calculate the free energy $F$ and find the
gate voltage dependence of the island charge difference $n(N,T)=\langle
\hat{n}\rangle= -\frac{1}{2 E_C}\frac{\partial F}{\partial N}+N$, where $\langle
\ldots \rangle$ denotes the thermodynamic average. The results are conveniently
expressed in terms of the rescaled temperature $\tilde T =T/E_J$, binding energy
$\tilde{\epsilon}_B=\epsilon_B/E_J$ and gate voltage $\tilde{N}$. Denoting the
zero temperature result by $ n(N,0) =n_0(\tilde{N})$,
\begin{equation}\label{eq:charge_zero}
n_0(\tilde{N})=\frac{\tilde{N}}{\sqrt{\tilde{N}^2 + 1}},
\end{equation}
we can express the island charge difference as
\begin{equation}\label{eq:charge}
 n(N,T)= \frac{2\sinh\left(\frac{\tilde{N}}{\tilde{T}}\right)
 e^{-|\tilde{N}|/\tilde{T}}+ n_0(\tilde{N})\,\, e^{\tilde{\epsilon}_B/\tilde{T}-\alpha}}{
 2\cosh\left(\frac{\tilde{N}}{\tilde{T}}\right)e^{-|\tilde{N}|/\tilde{T}}+
e^{\tilde{\epsilon}_B/\tilde{T}-\alpha}}.
\end{equation}
Here we introduced the notation $\alpha=\frac{1}{2}\ln{\frac{8\pi\Delta E_J
\tilde{T}}{\delta^2}}$.

\begin{figure}
\includegraphics[scale=0.8,bb=0 0 272 245]{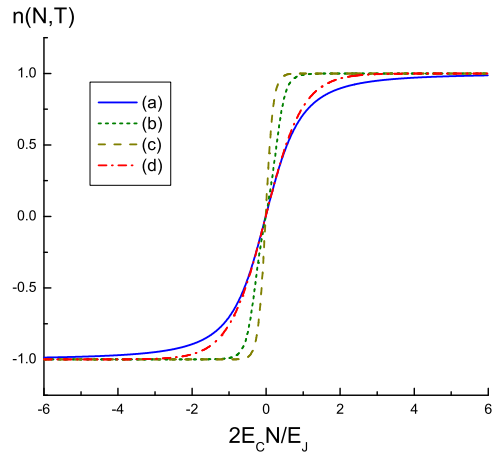}
\caption{The dependence of the island charge difference on the rescaled gate
voltage $2E_C N/E_J$ for temperatures $T=0$ (a), $T=0.08 E_J$ (b), $T=0.2E_J$
(c), $T=E_J$ (d). The temperature dependence of the step width is clearly
non-monotonic.} \label{step}
\end{figure}

The gate voltage dependence of the island charge is plotted in Fig.~\ref{step}
for several temperatures and typical experimental parameters~\cite{Bibow2002},
$\Delta\approx 2K$, $g\approx 1$, $\delta\approx 10^{-4}K$. Remarkably, the
temperature dependence of the step width is non-monotonic and shows a minimum at
the ionization temperature of the bound state $T_{i}\sim E_J /\alpha$.
Qualitatively, the non-monotonic temperature dependence of the step width can be
understood as follows. At $T < T_i$ the quasiparticle resides predominantly in
the bound state, and the width of the step is determined by the binding energy,
$\delta N \sim E_J/E_C$. At $T > T_i$ the quasiparticles resides in the
continuum, and the width of the step is determined by the temperature, $\delta N
\sim T/E_C$.  Since according to the Saha rule the ionization temperature $T_i$
is smaller than the binding energy $E_J$ the temperature dependence of the step
width is nonmonotonic.

We now turn to the case of an even number of electrons in the device, which we
take to be 2 to specific. For brevity we set $T=0$ from here on.

\begin{figure}
\includegraphics[scale=0.85,bb=0 0 262 231]{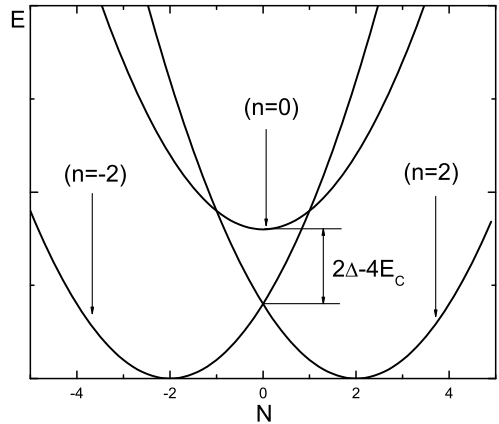}
\caption{Energy diagram for the Josephson energy calculation.
Electrostatic energies of the relevant charge states are plotted as a function of
$N$. The states with $n=\pm2$ are coupled by tunneling via the continuum of
states with $n=0$.} \label{states}
\end{figure}

First we study the dependence of the strength of the Josephson coupling between
the grains on $\Delta -2E_C$ for $\Delta> 2 E_C$. We consider the vicinity of the
degeneracy point ($|N|\ll 1$) between the two charge states, $n=\pm2$, with one
extra Cooper pair in the system. For $\Delta>2E_C$ the energies of these two
states lie below the bottom of the continuum of states with $n=0$ and one
quasiparticle in each island, see Fig~\ref{states}.

For $\Delta-2 E_C$ sufficiently large the Josephson energy can be obtained from
perturbation theory in tunneling~\cite{Matveev1993}:
 \begin{equation} \label{MatveevEJ}
\tilde
E_J=E_J\frac{4}{\pi^2}\int^{\infty}_{0}\frac{d\xi_k}{\varepsilon_k}\int^{\infty}_{0}
\frac{d\xi_p}{\varepsilon_p}\frac{\Delta}{(\varepsilon_k+\varepsilon_p-4E_C)},
\end{equation}
where $\varepsilon_{k(p)}=\sqrt{\xi^2_{k(p)}+\Delta^2}$.

At $\Delta \to 2E_C$ the integral in Eq.~(\ref{MatveevEJ}) diverges
logarithmically. This divergence arises from tunneling through the states of the
continuum with $\xi_{k(p)} \ll \Delta$. These states become strongly hybridized
with the two discrete states with $n=\pm 2$. To find the Josephson energy
with logarithmic accuracy we look for the wave function in the form of a
linear combination of the low energy states,
\begin{equation}
\label{psiJ}
|\Psi\rangle=\Psi_{20}|2,0\rangle+\Psi_{02}|0,2\rangle+\sum_{k,p,\sigma}
C_{p,k,\sigma}|k,\sigma ;p,-\sigma\rangle.
\end{equation}
Here $|2,0\rangle$ and $|0,2\rangle$ denote the states with $n=2$ and $n=-2$
respectively, and $|k,\sigma ;p,-\sigma\rangle$ denotes the $n=0$ state with
quasiparticles in state $k,\sigma$ in the left dot and $p,-\sigma$ in the right
one. Solving the Shroedinger equation with this wavefunction is straightforward.
The contunuum spectrum is unaffected by the presence of tunneling. The energies
$E$ of discrete spectrum can be found from the equation
\begin{equation}\label{eq:discrete}
  E =\frac{E_{20} +E_{02}}{2}
%    \left( \frac{1}{E-E_{20}}+ \frac{1}{E-E_{02}}\right)
    + \frac{\pi(E-E_{20})(E-E_{02})}{8 E_J \ln{\frac{E_{11} +2\Delta -E}{E_{11}
    +3\Delta-E}}}.
\end{equation}
Here $E_{11}=E_C N^2$   is the  electrostatic energy of the $|k,\sigma
;p,-\sigma\rangle$ states, and $E_{20(02)}=E_C(2\mp N)^2$ are those of the
$|2,0\rangle$ and $|0,2\rangle$ states respectively. For $\Delta > 2E_C$ this
equation has two solutions. The Josephson energy $\tilde E_J$ is given by the
energy splitting between these states at the degeneracy point $N=0$,
\begin{equation}\label{maxEJ}
\tilde E_J=\frac{8}{\pi}E_J\ln{\frac{\Delta}{\max \{E_J,\Delta-2E_C\}}},
\end{equation}
where $E_J$ is the Ambegaokar-Baratoff expression for the Josephson energy
defined below Eq.~(\ref{eq:gs}).

For $E_J < \Delta -2 E_C \ll \Delta$ this expression reproduces the logarithmic
asymptotic of the perturbative result (\ref{MatveevEJ}). For $ 0 < \Delta -2 E_C
\lesssim E_J $ perturbation theory breaks down and the Coulomb enhancement factor
of the Josephson energy saturates for small $g$ at the value
\begin{equation}\label{enhancement}
\frac{\tilde E_J}{E_J}=\frac{8}{\pi}\ln{\frac{8}{g}}.
\end{equation}

Equation (\ref{maxEJ}) shows that the Coulomb enhancement factor of the Josephson
energy in a double-island qubit can be much larger than that for a single grain
connected to a bulk lead~\cite{Matveev1993,Joyez1994}, which cannot exceed $1.3$.
This agrees with the recent experimental observation of an enhancement by a
factor of $\sim 3$ in double-island qubits~\cite{Bibow2002}.

For $\Delta -2 E_C <0$ the discrete level with the higher energy disappears as a
result of merging with the continuum, whereas the lower level continues to be
separated by a finite gap from the continuum and represents the ground state of
the system. The gate voltage dependence of its energy, which can in principle be
obtained by solving Eq.~(\ref{eq:discrete}), determines the island charge
difference at zero temperature, $n(N,0)$.

For $-E_J \lesssim \Delta -2 E_C <0$ the gap between the ground state and the
bottom of the continuum is of order $E_J$. The gate voltage dependence of the
ground state energy and the island charge remain qualitatively the same as those
for $\Delta -2 E_C >0$. In particular, the charge plateaux with $n=\pm 2$ are
joined by a smooth curve that lacks an intermediate plateau at $n=0$.

If $\Delta$ decreases even further, $ \Delta -2 E_C < -E_J$, the island charge
difference develops a plateau, $n(N,0)\approx 0$ for $|N| < 1-\Delta/2E_C$. In
this regime the ground state corresponds to a bound state of two quasiparticles,
one in each island, with the gap between the ground state and the bottom of the
continuum being exponentially small, $E_{11} +2\Delta -E \approx
\Delta\exp\left(-\frac{\pi (E_{20} -E_{11})(E_{02} -E_{11})}{4E_J(E_{20}+E_{20}
-2 E_{11})}\right)$. Near the charge degeneracy points $N_{\pm}=\pm
\left(1-\frac{\Delta}{2E_C}\right)$ between the states with $n=0$ and $n=\pm 2$
Eq.~(\ref{eq:discrete}) simplifies. As a result the island  charge difference can
be found from the implicit relation,
\begin{equation}\label{eq:charge_even}
      \frac{n}{2-n}+\ln{\frac{n}{ (2-n)}}=\frac{\pi N^*}{2} + \ln{\frac{2\pi }{g }}.
\end{equation}
Here we assumed $N$ to be positive and denoted by $N^*=\frac{2
E_C}{E_J}\left(N-1+\frac{\Delta}{2E_C}\right)$ a rescaled gate voltage distance
from the charge degeneracy point. For $N<0$ the island charge difference can be
obtained by noting that $n$ is an odd function of $N$. Away from the charge
degeneracy point, i.e. for $|N^*|\gg 1$, we obtain the approximate expressions
for the island charge difference: $n=\frac{4\pi}{g}\exp\left(-\frac{\pi
|N^*|}{2}\right)$ for $N^*<0$, and $ n=2-\frac{4}{\pi N^*}$ for $N^*>0$.

In summary, we have studied the low temperature thermodynamic properties of a
double-island qubit in the limit of the vanishing single particle mean level
spacing. We have shown that the ground state of the system is always separated
from the continuum of excited states by a finite gap, and determined the gate
voltage dependence of the island charge.

For an odd number of electrons in the device the ground state
corresponds to a bound state of the intrinsic quasiparticle. The
binding energy of this state is given by Eq.~(\ref{eq:gap}). It
attains its biggest value given by the Josephson energy $E_J$ at
the charge degeneracy point. The gate voltage dependence of the
island charge is given by Eqs.~(\ref{eq:charge_zero}) and
(\ref{eq:charge}). The presence of the bound state leads to a
non-monotonic temperature dependence of the width of the Coulomb
blockade step. The minimal width is achieved at the ionization
temperature of the bound quasiparticle state $T_i\sim
E_J/(\frac{1}{2}\ln{\frac{8\pi\Delta E_J}{\delta^2 }})$.

For an even number of electrons the gate voltage dependence of the ground state
energy and the island charge difference are determined by
Eqs.~(\ref{eq:discrete}) and (\ref{eq:charge_even}) respectively. For $\Delta <
2E_C$ and $|N|< 1-\Delta/2E_C$ the two intrinsic quasiparticles are bound to the
tunneling contact. The binding energy is exponentially small, see discussion
above Eq.~(\ref{eq:charge_even}). For $\Delta > 2E_C$ the Coulomb stimulation of
the Josephson energy in the double-island system can significantly exceed that
for a single grain coupled to a bulk electrode, see Eqs.~(\ref{maxEJ}) and
(\ref{enhancement}). This is in agreement with the experimental findings of
Ref.~\cite{Bibow2002}.

In deriving our results we neglected virtual transitions to higher energy states,
such as the non-resonant charge states and states with more than one
quasiparticle in the system. Such transitions lead to corrections that remain
small at the charge degeneracy point and amount to renormalization of the
charging energy of the system~\cite{LarkinOvchinnkov,Matveev1998}. Thus $E_C$
should be understood as the renormalized charging energy.

Our results can be directly tested by sensitive charge measurements techniques
that became available recently~\cite{Schoelkopf,Lehnert2003}. The bound
quasiparticle state is also present in a single superconducting grain coupled to
a superconducting lead when the electrostatic energy favors an odd number of
particles in the grain~\cite{Lafarge1991}. However in contrast to the latter
case~\cite{Matveev1998} the gate voltage dependence of the island charge in the
double-island qubit has no discontinuity at the charge degeneracy point. This
difference is due to the absence of thermal quasiparticles in the double-island
system at low temperatures.

This work was supported by the NSF Grant DMR-9984002 and by the David and Lucille
Packard Foundation. We are grateful to L.~I.~Glazman and K.~W.~Lehnert for
fruitful discussions.


\begin{thebibliography}{99}
\bibitem{footnote} This assumption is not true for the experiments
of Ref.~\cite{Bibow2002}, where the two islands were coupled to
two superconducting electrodes by highly resistive tunneling
contacts. In this setup the total number of electrons in the qubit
can be kept fixed by adjusting the pairing gap $\Delta$ and the
corresponding gate voltages appropriately.
\end{thebibliography}
\end{document}